\documentclass[twocolumn,showpacs,preprintnumbers,amsmath,amssymb,twoside]{revtex4}
\usepackage{graphicx}% Include figure files
\usepackage{dcolumn}% Align table columns on decimal point
\usepackage{bm}% bold math

\begin{document}
\newcommand{\NCOx}{Na$_{\mbox{\scriptsize x}}$CoO$_{\mbox{\scriptsize 2}}$}
\newcommand{\NCO}{Na$_{\mbox{\scriptsize 0.75}}$CoO$_{\mbox{\scriptsize 2}}$}
\newcommand{\tco}{T$_{\mbox{\scriptsize CO}}$}
\newcommand{\tso}{T$_{\mbox{\scriptsize SO}}$}
\newcommand{\tc}{T$_{\mbox{\scriptsize C}}$}
\newcommand{\tilt}{I$_{\mbox{\scriptsize 300}}$}
\newcommand{\gamm}{$1/\gamma^2$}
\newcommand{\LPSMO}{(La$_{\mbox{\scriptsize 1-y}}$Pr$_{\mbox{\scriptsize y}}$)$_{\mbox{\scriptsize 7/8}}$Sr$_{\mbox{\scriptsize 1/8}}$MnO$_{\mbox{\scriptsize 3}}$}

\newcommand{\quadru}{ $(1\pm 0.25,\pm 0.25,0)$}
\newcommand{\doub}{ $(1\pm 0.5,\pm 0.5,0)$}

%

%\begin{document}

\title{Stripe correlations in Na$_{0.75}$CoO$_2$}

\author{J. Geck$^1$}
\author{M. v. Zimmermann$^{2}$}
\author{H. Berger$^{3}$}
%\author{J. Fink$^1$}
\author{S.V. Borisenko$^1$}
\author{H. Eschrig$^1$}
\author{K. Koepernik$^1$}
\author{M. Knupfer$^1$}
\author{B. B\"uchner$^1$}
% \thanks is optional - remove next line if not needed
%\thanks{\emph{Present address:} Insert the address here if needed}
%}                     % Do not remove
%
%\offprints{}          % Insert a name or remove this line
%

\affiliation{$^1$Leibniz Institute for Solid State and Materials Research  IFW Dresden, Helmholtzstr. 20, 01069 Dresden, Germany}

\affiliation{$^2$Hamburger Synchrotronstrahlungslabor HASYLAB at
 Deutsches Elektronen-Synchrotron DESY, Notkestr. 85, 22603 Hamburg,
 Germany}
\affiliation{$^3$Institut de physique de la mati\'{e}re complex (IPMC),EPF Lausannne, 1015 Lausanne, Switzerland}

\date{Received: \today}
% / Revised version: date}
% The correct dates will be entered by Springer
%
\begin{abstract}

We unambiguously demonstrate, based on high-energy x-ray diffraction data and LDA calculations, that sodium-density stripes are formed in \NCO\/
at low temperatures and rule out the previously proposed Na-ordering models. The LDA calculations prove, that the sodium-density stripes lead to
a sizeable dip in the density of the Co-states at the Fermi level, pointing to band structure effects as a driving force for the stripe
formation. This indicates that the sodium ordering is connected to stripe-like charge correlations within the CoO$_2$ layers, leading to an
astonishing similarity between the doped cuprates and the \NCOx\/ compounds.

\end{abstract}

\pacs{71.30.+h, 61.10.Eq, 64.60.Cn, 71.27.+a}
                              %display desired
\maketitle

The many-body effects in correlated electron systems based on square lattices often result in the development of intrinsic charge
inhomogeneities \cite{OrensteinScience00,TokuraScience00,TokuraPT03}. A particular prominent example is the so-called stripe ordered phase in
the cuprates, where the doped holes condense into charge stripes within the physically relevant CuO$_2$-planes \cite{OrensteinScience00}. Up to
date, this stripe order is discussed intensively in relation the high-temperature superconductivity and the anomalous metallic state of these
materials.

Recently \NCOx , being a correlated electron system based on a triangular lattice, has attracted considerable attention, due to the discovery of
superconductivity  in water-intercalated compounds \cite{TakadaNature03} as well as the outstanding physics of the non-hydrated materials
\cite{FooPRL04}. More specifically, for the non-hydrated \NCOx\/ materials with $x>0.5$ an anomalous metallic state with a thermopower ten times
larger than that of a typical metal, the coexistence of localized moments and electron itinerancy as well as unusual charge and spin order
phenomena have been reported \cite{WangNature03,GavilanoPRB04,BernhardPRL04}. These extraordinary electronic properties cannot be reconciled
with any conventional concept and remain to be a challenge for both theory and experiment.

This raises the question whether the outstanding physical properties of \NCOx\/ with $x\sim 0.7$ are also related to the presence of charge
inhomogeneities, which are formed within the CoO$_2$-planes in this case. To clarify this question, in the first step of this study, we
performed a detailed characterization of structural modulations in \NCO\/ by means of high-energy x-ray diffraction.
As has been proven by numerous previous experiments, HXS provides a highly sensitive probe for the extremely weak structural modulations related
to the formation of charge inhomogeneities \cite{NiemoellerEPJB99,vZimmermannEPL98,VigliantePRB97,GeckPRB02}. At the used photon energies around
100\,keV, the x-ray penetration depth is of the order of 1\,mm, which renders this technique insensitive to surface effects and guarantees the
detection of bulk properties.

For the present experiment, which was conducted at the beamline BW5 of the HASYLAB in Hamburg, the photon energy was set to 99.79\,keV
($\lambda=0.124$\,$\rm \AA$) and the longitudinal and transverse resolution at the (100) position was set to  0.009\,\AA$^{-1}$ (FWHM) and
0.015\,\AA$^{-1}$ (FWHM), respectively. Further details about the experimental setup can be found in the literature \cite{BouchardSRN98}.

\begin{figure}[t!]
\center{
\resizebox{0.99\columnwidth}{!}{%
   \includegraphics[clip, angle=0]{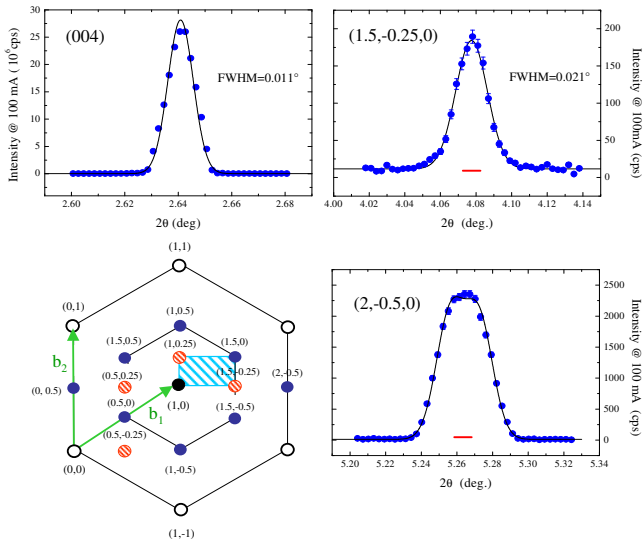} }}

\caption{ (color online) Comparison between different radial ($\Theta/ 2\Theta$-) scans taken at T=8\,K.   The solid lines are fits to the data (see text) and % with a FWHM that is given in the plots.
the bar below the superstructure peaks represents the experimental resolution. The observed superstructure peaks around (100) are summarized in
the lower left panel, where the hatched (light blue) rectangle indicates the unit cell of the reciprocal lattice.
Open, solid (blue) and hatched (red) circles indicate Bragg reflections, superstructure peaks observed for all investigated temperatures, and
superstructure peaks only observable below \tso, respectively.} \label{fig:1}
\end{figure}

The high-quality  \NCO\/ single crystals used for the x-ray diffraction studies were grown by the sodium chloride flux methods as thoroughly
described elsewhere \cite{Iljev04}. The mosaicity spread of the sample was determined to be about 0.24\,$^{\circ}$. Furthermore, the observed
room temperature lattice parameters  $a_H=2.83$\,\AA\/ and $c_H=10.84$\,\AA \/ of the \NCO \/ sample under study agree well with the values
reported in the literature \cite{HuangPRB04b}. Since the $c$-axis depends strongly on $x$ \cite{HuangPRB04a,FooPRL04}, the resolution limited
radial ($\Theta/ 2\Theta$-) scan through the (004) reflection at T=8\,K shown in Fig.\,\ref{fig:1} verifies a well-defined and homogeneous
sodium concentration in the probed sample volume.

\begin{figure}[t!]
\center{
\resizebox{0.75\columnwidth}{!}{%
   \includegraphics[clip, angle=0]{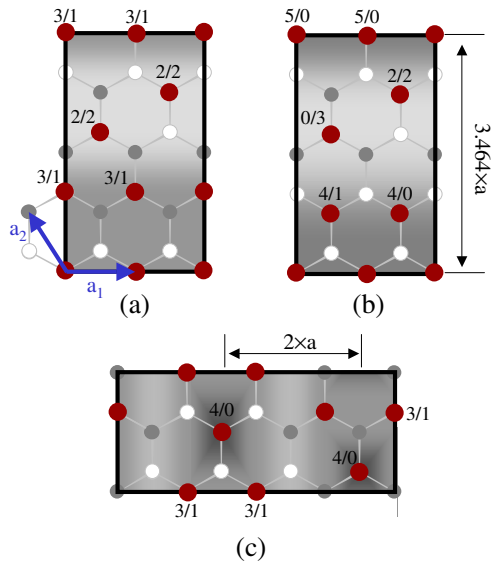} }}

\caption{(color online) Models for the sodium superlattice in \NCO.  Projections of superstructure unit cells on the
$\left(\mathbf{a}_{1},\mathbf{a}_{2}\right)$-plane are shown. Small gray and white circles indicate Na$\left(1\right)$- and
Na$\left(2\right)$-sites, respectively. The occupied Na-sites are indicated by big solid (red) circles. Structures (a) and (b) are compatible
with experimental diffraction peaks; (a) with lowest LDA total energy. (c) is the superstructure reported in the literature
\cite{ZandbergenPRB04a,zhang05}, which is excluded by the experimental diffraction pattern. The numbers $m/n$ indicate the number of
first/second Na-Na neighbors and the corresponding sodium-density variations are indicated by the gray shadings.} \label{fig:4}
\end{figure}

During a survey in reciprocal space at 8\,K,  a number of superstructure reflections were observed around the (100) reflection in the
(HK0)-zone, which are summarized in Fig.\,\ref{fig:1}. In this figure, the indicated reciprocal lattice vectors $\mathbf{b}_{1,2}$ correspond to
the direct lattice vectors $\mathbf{a}_{1,2}$  that are parallel to the Na- and CoO$_2$-layers (cf. Fig.\,\ref{fig:4}). Corresponding
superstructures were also observed around the (110) position. Radial scans through the superstructure reflections at the commensurate
(1.5,-0.25,0) and (2,-0.5,0) positions taken at T=8\,K are shown on the right hand side of Fig.\,\ref{fig:1}. The observation of superstructure
reflections at equivalent positions together with the fact that these reflections do not vanish upon a rotation around the scattering vector,
excludes multiple scattering as a possible origin for the observed intensities.
The intensity profile at the (1.5,-0.25,0) position was fitted by a single Lorentzian squared function. The FWHM determined by this fit together
with the experimental resolution determined at the (100) position yield a correlation length of about 300\,$\rm \AA$ for the corresponding
superstructure modulation.

\begin{figure}[t!]
\center{
\resizebox{0.75\columnwidth}{!}{%
   \includegraphics[clip, angle=0]{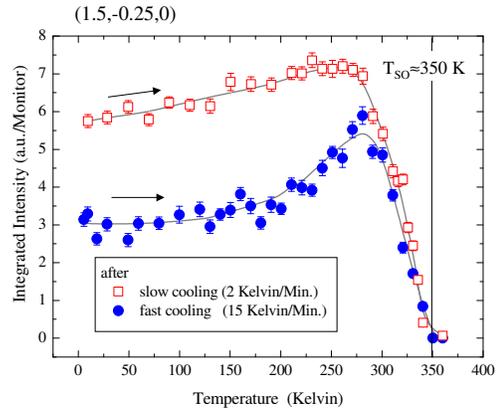} }}

\caption{(color online) Temperature dependence of the integrated intensity at the (1.5,-0.25,0) position. After cooling down to low temperatures
at different cooling rates, both measurements have been performed at increasing temperature. The solid lines are guides to the eye.}
\label{fig:2}
\end{figure}

In order to obtain information about the origin of the observed superstructures, we performed temperature dependent measurements at two
representative positions in reciprocal space: In Fig.\,\ref{fig:2}, the temperature dependence of the integrated intensity at the (1.5,-0.25,0)
position is displayed, showing that this reflection appears only below \tso=350\,K.
In addition to this, the intensity of the (1.5,-0.25,0) reflection depends strongly on the cooling process, as demonstrated by the two
measurements shown in Fig.\,\ref{fig:2}. The first measurement was performed with increasing temperature after the sample had been cooled down
slowly at 2\,K/min (open symbols). The second measurement was also taken upon heating, but this time the sample had been cooled down rapidly at
about 15\,K/min (closed symbols). The large differences between the two measurements are evident.
This is a strong indication for an ordering phenomenon that is related to the ordering of sodium ions, where relatively slow ion-hopping
processes are involved. Upon rapid cooling, the sodium order is frozen, i.e. it cannot develop completely, whereas it can be established to a
higher extend if the sample is cooled down slowly. Upon increasing the temperature again,  the thermally activated sodium ions become more and
more mobile, which allows to remove defects in the sodium superlattice  before the order finally vanishes upon heating at \tso.
The conclusion that the superlattice modulation below \tso \/ is not primarily related to charge order within the CoO$_2$ but to sodium order is
further supported by the metallic resistivity below 350\,K of \NCOx\/ compounds around x=0.75 \cite{FooPRL04}.

\begin{figure}[t!]
\center{
\resizebox{0.75\columnwidth}{!}{%
   \includegraphics[clip, angle=0]{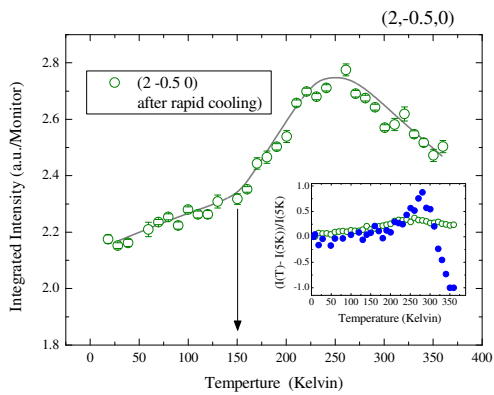} }
}

\caption{(color online) Temperature dependence of the integrated intensity at the (2,-0.5,0) position (open symbols). The measurement has been
performed with increasing temperature after rapid cooling. The solid line is just a guide to the eye. A comparison of the relative intensity
changes for the (1.5,-0.25,0) (full symbols) and the (2,-0.5,0) reflection is given in the inset.} \label{fig:3}
\end{figure}

As a second reflection, we have studied the (2,-0.5,0) peak and its  temperature dependence. The integrated intensity of this reflection
measured after rapid cooling is shown in Fig.\,\ref{fig:3}. The measurement after rapid cooling shown in Fig.\,\ref{fig:2} and the data given in
Fig.\,\ref{fig:3} have been obtained during the same run, verifying a completely different temperature dependence of the two reflections. As can
be observed in Fig.\,\ref{fig:3}, the (2,-0.5,0) reflection does not disappear above 350 K, but is observed in the whole investigated
temperature range instead. The slope change of the (2,-0.5,0) intensity around 150 K roughly coincides with the temperature regime where the
(1.5,-0.25,0) intensity starts to increase considerably. Furthermore, the intensity at the (2,-0.5,0) reflection also varies across the phase
transition at 350 K, indicating that these two superlattice modulations are coupled. Nonetheless, the temperature dependent variation of the
(2,-0.5,0) reflection is relatively weak, as illustrated in the inset of Fig.\,\ref{fig:3}.
To conclude so far,  %temperature dependencies of the (2,-0.5,0) and the (1.5,-0.25,0) reflection show that
the low temperature superstructure is established in two steps: First, the doubling of the unit cell  signaled by the (2,-0.5,0) reflection
occurs at temperatures well above 360\,K and then, the structural modulation related to the (1.5,-0.25,0) peak appears below \tso=350\,K, which
can clearly be attributed to sodium order.

The temperature dependent measurements described above agree well with a recent neutron powder diffraction study on Na$_{0.75}$CoO$_2$
\cite{HuangPRB04b}, where  a structural phase transition has been observed in the very same temperature regime and has been discussed in terms
of a sodium rearrangement. However, in this study only a symmetry change related to the Na-sites has been revealed, whereas no enlargement of
the unit cell has been reported.

Based on the experimental data for T=8\,K we can unambiguously identify the unit cell of \NCO\/ at low temperatures: We start from the hexagonal
layered lattice structure belonging to the $P6_3/mmc$ space group \cite{HuangPRB04b}. In this setting, the CoO$_2$-layers and the Na-planes are
parallel to the $(\mathbf{a}_1,\mathbf{a}_2)$-planes and alternate along the perpendicular stacking direction $\mathbf{c}$.
The lattice symmetry allows for two in-equivalent lattice sites within the Na-planes, referred to as Na(1) and Na(2) in the following, which
together form a honeycomb lattice (cf. Fig.\,\ref{fig:4}). However, not all the sites of this lattice can be occupied, because the Na(1)-Na(2)
nearest neighbor distance is far too short to allow a simultaneous occupation.
If $\mathbf{b}_{1}$, $\mathbf{b}_{2}$ are the reciprocal lattice vectors of the $P6_3/mmc$ lattice, then the additionally found superstructure
peaks at low temperatures correspond to an orthogonal basis $\mathbf{b}_{1}/2-\mathbf{b}_{2}/4$, $\mathbf{b}_{2}/4$ (cf. Fig.\,\ref{fig:1}).
Since we did not observe any additional structural modulation along $\mathbf{c}$, the HXS data implies an orthorhombic unit cell of the direct
lattice, spanned by $2\mathbf{a}_{1}$, $2\mathbf{a}_{1}+4\mathbf{a}_{2}$, and $\mathbf{c}$ (cf. Fig.\,\ref{fig:4}\,$(a)$ and $(b)$).

In the second step of this study, we performed LDA calculations \cite{perdewwang92} based on the identified supercell using the FPLO package
version 5.20 \cite{koe99}. The calculations were performed non-spin-polarized and all calculations were done for a fixed unit cell volume. To
compare to previous calculations in the literature \cite{zhang05}, we used $a_{H}=2.82\textrm{Å}$ and $c_H=10.89\textrm{Å}$, which slightly
differ from the observed values by about 0.5\,\%.
The aim of these calculations was twofold: (i) To determine the lowest energy state in agreement with the HXS data, since  several sodium
ordering patterns are possible within the observed supercell as demonstrated by Fig.\,\ref{fig:4}\,(a) and (b). (ii) To determine the effect of
the sodium order on the electronic states of the CoO$_2$-planes. %

The free parameter of the oxygen Wyckoff position was relaxed for each structure and resulted in the same oxygen position for all Na patterns
considered. In order to assess our calculations we recalculated the total energies of a number of structures considered in Ref. \cite{zhang05}
and obtained very good agreement in the structural energy differences. The LDA calculations yield the lowest total energy for the structure
given in Fig.\,\ref{fig:4}\,(a), which amounts to $E_{a}=1662.091$ mHartree per Na$_{0.75}$CoO$_{2}$ formula unit.
The two structures shown in Fig.\,\ref{fig:4}\,(b) and (c) correspond to higher total energies. We obtain $E_{b}-E_{a}=1.3$ mHartree and
$E_{c}-E_{a}=1.7$ mHartree, respectively.
With respect to the stacking in the third direction, two possibilities were considered with an inversion center at the Co site and midway
between two nearest neighbor Co sites, respectively. However, the corresponding energy differences are negligible. % on a mHartree scale.
The differences between the structures (\emph{a}), (\emph{b}) and (\emph{c}) can essentially be understood by Coulomb energy arguments:
(\emph{a}) and (\emph{c}) comprise a charge density wave of comparable amplitude, but with wave vector $2\mathbf{a}_{1}+4\mathbf{a}_{2}$ in the
case (\emph{a}), while (\emph{c}) has the shorter wave vector $2\mathbf{a}_{1}$. The structure (\emph{b}) has also a charge density wave with
wave vector $2\mathbf{a}_{1}+4\mathbf{a}_{2}$, however with a considerably larger amplitude.

Note, that the model reproduced in Fig.\,\ref{fig:4}\,(c) corresponds to the lowest energy state reported by Zhang {\it et al.} based on
electron diffraction (ED) data \cite{ZandbergenPRB04a,zhang05}. Obviously, the supercell determined by ED disagrees with the presented HXS
results.  This discrepancy is most likely related to the different sample volumes, which are probed by the two techniques. HXS, being a bulk
probe, is not sensitive to the sample surface, while ED is surface sensitive. This implies that the sodium ordering is strongly effected by
surface effects, which should be taken into account for the interpretation of surface sensitive probes like angular resolved photoemission.

The obtained ground state corresponds to the formation of sodium-density stripes within the Na-planes, as indicated by the gray shading in
Fig.\,\ref{fig:4}\,(a). The LDA calculations prove, that this ordering produces a sizeable dip in the density of the Co-states at the Fermi
level, pointing to band structure effects as a driving force for the stripe formation.
This is a very strong indication that the ordering within the Na-planes is coupled to intrinsic stripe-like charge correlations within the
CoO$_2$-planes.

%-----------------------------------------------------------------------------------------

Taking into account that the hydrated \NCOx\/ samples have been shown to be of the composition Na$_{0.337}$(H$_3$O)$_{z}$CoO$_2\cdot y$H$_2$O,
where the additionally intercalated H$^+$-ions of the H$_3$O$^+$ also occupy Na-sites \cite{MilnePRB04}, it follows that the effective doping of
the superconducting materials corresponds to \NCOx\/ with $x\approx0.7$. This leads to an astonishing analogy to the high-temperature
superconducting cuprates: In the case of the doped cuprates, the cooperative octahedral tilts in the so-called LTT-phase lead to a pinning
potential for charge stripes that can become static, if the tilt angle is large enough.
In the \NCOx\/ materials, the electric field that is caused by the sodium-density stripes might induce a pinning potential for charge stripes
within the adjacent CoO$_2$ layers as well. The electric field in the CoO$_2$ layers caused by the sodium order depends on the distance between
the Na- and the CoO$_2$ layers; i.e. it can be tuned by the $c_H/a_H$-ratio. One may speculate, whether the $c_H/a_H$-ratio in the \NCOx\/
system plays a similar role as the tilt angle in the LTT phase of the doped cuprates.
%
%In the case of the \NCOx, system the electric field caused by the sodium-vacancy stripes might result in a pinning potential that can be
%tuned by the $c/a$-ratio.
%
In terms of this scenario, one would expect a tendency to develop static stripe order for a small $c_H/a_H$-ratio (no water intercalation),
which become dynamic upon increasing $c_H/a_H$ by water intercalation. This would naturally explain the appearance of superconductivity upon
water intercalation. However, this scenario has certainly to be verified by further experiments which are capable to detect directly charge
order within the CoO$_2$ layers.

In conclusion, combining experimental HXS data and LDA model calculations we have unambiguously identified sodium-density stripe order in \NCO ,
which has a strong impact on the electronic states of the CoO$_2$-planes. The calculations point to band structure effects as a driving force
and indicate that intrinsic charge-stripe correlations are present within the two-dimensional triangular CoO$_2$-layers.

{\bf Acknowledgements:} We would like to thank Dr. Bussy (University of Lausanne) for the micro probe analysis. This work was supported by the
Swiss NCCR research pool MaNEP of the Swiss NSF and the DFG with the SFB 463 and the FOR 538 research units.

\bibliographystyle{apsrev}

\end{document}